\documentclass[aps,twocolumn,showpacs,preprintnumbers,amsmath,amssymb,floatfix]{revtex4} 
\usepackage[]{graphicx}
\usepackage{amsmath}

\begin{document} 

\title{Nonlinear excitations in DNA: Aperiodic models versus actual 
genome sequences}

\author{Sara Cuenda}
\email{scuenda@math.uc3m.es}

\affiliation{%
Grupo Interdisciplinar de Sistemas Complejos (GISC) and
Departamento de Matem\'aticas\\
Universidad Carlos III de Madrid, Avenida de la Universidad 30, 28911
Legan\'es, Madrid, Spain}%
\homepage{http://gisc.uc3m.es}

\author{Angel S\'anchez}
\email{anxo@math.uc3m.es}

\affiliation{%
Grupo Interdisciplinar de Sistemas Complejos (GISC) and
Departamento de Matem\'aticas\\
Universidad Carlos III de Madrid, Avenida de la Universidad 30, 28911
Legan\'es, Madrid, Spain, and\\
Instituto de Biocomputaci\'on y F\'\i sica de Sistemas Complejos, 
Universidad de Zaragoza, 50009 Zaragoza, Spain
}%

\date{\today}
\begin{abstract}
We study the effects of the genetic sequence on the propagation of nonlinear excitations
in simple models of DNA in which we incorporate
actual data from the human genome. 
We show that kink propagation requires forces over a certain 
threshold, a phenomenon already found for aperiodic sequences
[F.\ Dom\'\i nguez-Adame {\em et al.}, Phys.\ Rev.\ E {\bf 52}, 2183 (1995)]. 
For forces below threshold, the final stop positions are
highly dependent on the specific sequence.
Contrary to the conjecture advanced by Dom\'\i nguez-Adame and coworkers,
we find no evidence supporting the dependence of the kink dynamics on 
the information content of the genetic sequences considered. 
We discuss possible reasons for that result as well as its practical 
consequences. Physically, 
the results of our model are consistent with the stick-slip dynamics
of the unzipping process observed in experiments. 
We also show that the effective potential, a collective 
coordinate formalism introduced by 
Salerno and Kivshar [Phys.\ Lett.\ A {\bf 193}, 263 (1994)] is a useful tool to
identify key regions in DNA that control the dynamical behavior of
large segments. 
As a side result, we extend the previous studies on aperiodic sequences by
analyzing the effect of the initial position of the kink, leading 
to further insight on the phenomenology observed in 
such systems.
\end{abstract}

\maketitle

\section{Introduction}

Nonlinear models are becoming widely used to capture and 
understand emergent phenomena in complex systems 
\cite{Scott}. The pioneering field in this direction was physics, 
where nonlinear phenomena have been the subject of intense study 
since the seminal work of Fermi, Pasta and Ulam 
almost fifty years ago \cite{FPU}. The success of this approach in
modelling of complex systems led scientists in other
fields to apply it to their own research. That is the case in
biology, where nonlinear models
were successfully introduced more than 20 years ago \cite{Davydov} 
(see also \cite{Davydov2} for references). 
One of the subjects where nonlinear models have been 
more productive is the modelling of DNA physics
\cite{Yak1,Gaeta1,Yak2,Gaeta2,ReviewMichel}, which began in 1980 
with the model proposed by 
Englander and coworkers \cite{Englander}. Since then, a lot of work has been devoted to
nonlinear excitations in DNA, both from the dynamics and the statistical
mechanics viewpoints. We refer the 
reader to \cite{Yak2,ReviewMichel} for historical accounts and 
extensive summaries of the available results, among which we want
to highlight the 
Peyrard-Bishop model \cite{PB} (see also the generalizations and
improvements proposed in \cite{DPB,DP}).  
In spite of its simplicity, this nonlinear lattice system can accurately 
(sometimes even quantitatively) reproduce the phenomenology experimentally
observed in DNA (see, e.g., \cite{Campa}).

Most of the
research done in the framework of the above mentioned models
refers to {\em homopolymers}, 
i.e., homogeneous DNA molecules consisting entirely of A-T or C-G base
pairs.
One problem 
of special interest was 
the thermal denaturation transition, that takes 
place at temperatures around 90$^{\circ}$ C when the two strands of the
DNA molecule separate. Having understood this transition quite well 
\cite{ReviewMichel,Wartell}, the main focus of research 
has turned dynamical and transport features. Additional motivation 
for such a shift arises from 
the capability to carry out experiments on single 
molecules, achieved in the last few years 
\cite{Nature}.
Another particularly important, related question is
genomics, or gene identification, which is profiting 
enormously
from the information obtained from dynamical models \cite{Yeramian}. 
Of course, these applications require unavoidably the study of 
DNA {\em heteropolymers}, where the distribution of A-T and C-G base 
pairs follows non-uniform, non-homogeneous sequences obtained from
genome analysis. In this context,
physical models are also being used to idenfity dynamically relevant
sites, such as promoters \cite{LosAlamos} in short genomes (e.g.,
viral). Further advance
along this line requires good models for DNA dynamics that 
can be computationally efficient, in order to treat much longer
sequences. Such models will also be of use to achieve a better
understanding the relationship between sequence, physical properties,
and biological function.

In this work, we address these issues in the framework of 
a very simple model, namely the Englander model \cite{Englander}, which 
we generalize by incorporating actual genetic sequences.
Our aim is to assess whether the so modified
model, in spite of its simplicity, reproduces the important features of  
DNA molecule dynamics accurately enough to be useful for genomic and 
related applications. We will also discuss an approximation in terms 
of an effective potential and its application to understanding the 
dynamics of nonlinear excitations in DNA. In addition, we will
test a conjecture, put 
forward by Dom\'\i nguez-Adame and coworkers \cite{yo}, that the 
dynamics of nonlinear excitations along DNA molecules should, at the
level of the Englander model, depend on the information content of
the chain. 
To this end, the paper is organized as follows. In Sec.\ \ref{sec:bac},
we review the Englander model and the previous attempts to include 
inhomogeneity or genetic information in it. 
the genetic sequence in it. In so doing, we will discuss in some detail
the results of Dom\'\i nguez-Adame {\em et al.}, extending their work
to consider the effects of changing the initial location of the 
propagating kinks. 
Subsequently, in Sec.\ \ref{sec:eff}, we 
present our results on the Englander model with human genome 
sequences. We analyze the results by comparing them to the phenomenology
observed on aperiodic sequences and to the experimentally observed 
DNA dynamics. Finally, Section \ref{sec:con} concludes the paper by 
summarizing our main results and their possible implications. 

\section{Background}
\label{sec:bac}

\subsection{Model definition: the homogeneous case}

\begin{figure}
\begin{center}
\includegraphics[height=2cm]{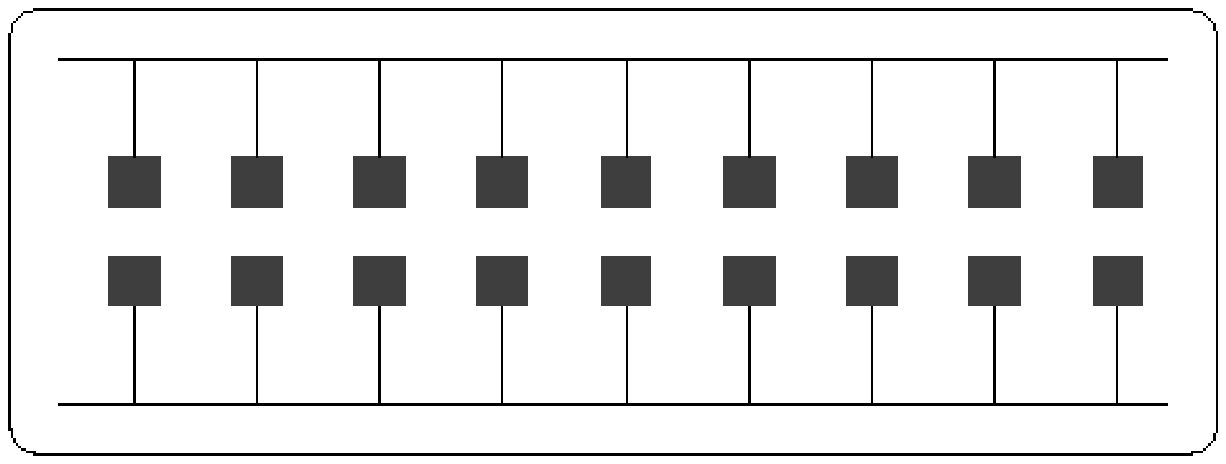}
\includegraphics[height=2cm]{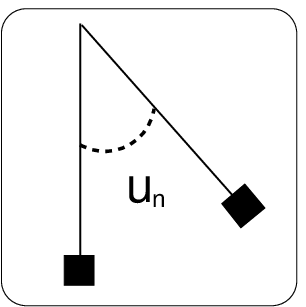}
\caption{\label{fig:modelo}
Sketch of the Englander model. The two sugar-phosphate backbones
of the DNA strands are depicted as 
two straight lines connected by the base pairs (bases are the
squares). The lower strand is assumed to be fixed and as
shown on the right, the angle $u_n$ is the deviation of the
upper base of the $n$-th pair with respect to the 
plane defined by the fixed base pairs in the lower strand.
}
\end{center}
\end{figure}

The model proposed by Englander {\em et al.}\ \cite{Englander} is 
sketched in Fig.\ \ref{fig:modelo}, 
where the dynamics of one of the strands of the 
DNA is represented as a chain of pendula; leaving the other strand fixed (the one at
the bottom in the picture), the base pairs of the (upper) strand behave 
like pendula in an effective
"gravitational field" caused by the tendency of the base pairs of the two 
strands to be facing each other. It must be realized that 
this model describes only the
dynamics of the base pairs around the sugar-phosphate backbone, which  
is assumed to be fixed as well. 
Introducing damping and an external force these pendula are described by the
discrete, dc driven, damped sine-Gordon model, given by
\begin{equation}
\label{1}
\ddot{u}_n-\frac{1}{a^2}(u_{n+1}-2u_n+u_{n-1})+q_n\sin u_n=-\alpha\dot{u}_n-F,
\end{equation}
where $a$ is the lattice spacing, $q_n$ is a site dependent constant which 
arises from the specific parameters for the pendulum at site $n$, $\alpha$
is the damping coefficient, and $F$ is a driving term possibly acting on 
the chain. We note at this point that Eq.\ (\ref{1}) reflects mainly
the difference in interactions between A-T and C-G base pairs, but
not other differences such as in mass or moments of inertia. It should
also be clear that many other relevant factors (geometry, inhomogeneous 
stacking interactions \cite{understanding})
have also been neglected, in the spirit of designing
a model as simple as possible that still behaves like DNA molecules.

\begin{figure}
\begin{center}
\includegraphics[width=7cm]{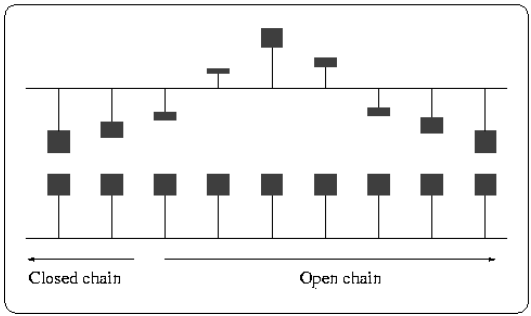}
\caption{\label{fig:kink_modelo}
Kink soliton in the sine-Gordon model. The kink joins a sector of the 
chain where bases are closed, $u_n=0$, to another one where bases have
performed a complete turn, $u_n=2\pi$. In this last part the chain is 
said to be open if the model is to represent mechanical denaturation.
}
\end{center}
\end{figure}

When $q_n=q$ (i.e., we have a homopolymer) in Eq.\ (\ref{1}) and the lattice 
spacing is very small, the system of ordinary differential equations (\ref{1})
can be very well approximated by its continuum limit. Letting
$u_n(t){\longrightarrow} u(x,t)$ when $a^2\ll q$, we find the driven, 
damped sine-Gordon equation:
\begin{equation}
\label{2}
\partial^2_t u-\partial^2_x u+q\sin u=-\alpha\partial_t u-F.
\end{equation}
It is well known \cite{Scott} that, in the absence of force, i.e., $F=0$,
Eq.\ (\ref{2}) possesses soliton solutions 
of the kink type, whose expression is 
\begin{equation}
\label{3}
\phi_{\pm}(x,t)=4\arctan\left\{\exp\left[\pm\sqrt{q}\left(
\frac{x-x_0-vt}{\sqrt{1-v^2}}\right)\right]\right\}
\end{equation}
where the plus or minus sign stands for kinks or antikinks respectively,
and $0\leq v\leq 1$ represents their velocity. In the case $F=0$ and
in the presence of damping, $\alpha\neq 0$,
the only possible value for the velocity is $v=0$.
When both damping and force are present, the balance between the 
two effects leads to kinks propagating at a constant, nonzero velocity.
An analytical expression for that velocity can be easily derived
[again, for homogenous sequences, i.e., from Eq.\ (\ref{2})]
from energy conservation arguments (see \cite{McL}):
\begin{equation}
\label{vel}
v=\left[1+q\left(\frac{4\alpha}{\pi F}\right)^2\right]^{-1/2}.
\end{equation}
Figure \ref{fig:kink_modelo} depictes the physical meaning of a kink 
solution in the context of DNA modelling: the bases of the upper chain
perform a complete, smooth turn around the sugar-phosphate backbone, from
$u_n=0$ to $u_n=2\pi$.

One feature of the model that deserves some comment is its mesoscopic 
character. Experiments and a great part of the theoretical
work done so far on mechanical denaturation or unzipping 
(\cite{Nature}; for the first unzipping experiments see \cite{Heslot})
deals with forces acting on the last 
base of one of the strands pulling them apart. In the 
perspective of Englander's model, we are not trying to describe
accurately the way the chain is pulled but rather we are looking 
at how this pulling affects the ``propagation'' of the opening. 
Of course, both approaches are different and need not yield the
same results, although one expects in principle that the qualitative
behavior should be the same. As we will see below, the advantage of
our approach is that it allows some analytic insight on the problem, 
much less amenable from the microscopic viewpoint. 

\subsection{Inclusion of inhomogeneities}

The results we have summarized in the previous subsection refer
mostly to the homopolymer case, which was the one initially 
considered by Englander {\em et al.}\ \cite{Englander}, 
As stated in the introduction, our main is to understand the 
effects of inhomogeneity, and particularly of genetic 
sequences obtained from real data. We stress that including 
inhomoengeity, i.e., $q_n$ depending on $n$ in Eq.\ (\ref{1}),
renders the theoretical approach in the preceding subsection 
unapplicable. To our knowledge, 
DNA sequences from living organisms were first studied in 
the framework of the Englander model by Salerno, who
considered the 
effects of the sequence by studying spontaneously travelling kinks 
along the T7A$_1$ promoter (a region of the genome preceding a gene where
the transcription activity starts) of the bacteriophage T7. Salerno
studied further other regions of the bacteriophage \cite{Salerno2},
finding that the dynamical activity of kinks should be very special
in the promoters. Recently, the problem was revisited in 
\cite{copia} following the sequencing of the whole genome of the
bacteriophage T7. The main conclusion of this work was that there 
indeed is a significantly higher degree of activity in promoter 
regions, even in the presence of noise. 

The results of the simulations reported in 
\cite{Salerno,Salerno2} were analyzed theoretically by 
Salerno and Kivshar \cite{Salkiv}, who, using a collective
coordinate approach (see, e.g., \cite{SB} for a review on collective 
coordinate techniques for soliton-bearing equations),
developed a description of the
kink dynamics in terms of an effective potential. We will not review in 
detail the procedure to obtain this effective potential but, instead, 
we will simply outline the main steps. The idea is to consider the
undamped model, i.e., Eq.\ (\ref{1}) with $\alpha=0$, on a finite
lattice of $N$ sites, described by the Hamiltonian
\begin{align}
\label{4}
H^{sG}[\{u\}]=\sum_{n=1}^N\{&\frac{1}{2}\dot{u}_n^2+
\frac{1}{2a^2}[u_{n+1}-u_n]^2\nonumber\\
&+q_n(1-\cos u_n)+Fu_n\}. 
\end{align}
We now take the following {\em Ansatz}
\begin{equation}
\label{ansatz}
\phi_n(X(t))=4\arctan\left\{\exp\left[\sqrt{q_{avg}}\left(
na-X(t)\right)\right]\right\},
\end{equation}
where $q_{avg}$ is the average value of $q_n$ over the chain, 
and $X(t)$ is a collective coordinate
variable which stands for the center of the kink. Substituting this
expresion into the Hamiltonian (\ref{4}), we arrive at 
\begin{equation}
\label{5}
E_{eff}=\frac{1}{2}\dot{X}^2+V_{eff}(n, \{q_n\}),
\end{equation}
which is formally equivalent to the energy of a particle subjected to 
the action of the effective potential $V_{eff}$. The exact formula for 
this potential is quite cumbersome and we do not need to reproduce 
it here. We refer the reader interested in the details of the calculation
and the full result to \cite{Salkiv,yo,copia2}.

\subsection{Aperiodic chains}\label{sec}

Since the sequences considered by Salerno \cite{Salerno,Salerno2} are
very small, the question arises immediately as to the real relevance
of those results. On the other hand, in 1995 there were not many 
genomic data available, making the question difficult to answer. 
To overcome this difficulty, 
Dom\'\i nguez-Adame and coworkers \cite{yo} proposed to mimic 
the behavior of DNA heteropolymers of biological relevance by 
replacing the sequence dependent values $q_n$ by an aperiodic, 
but fully deterministic sequence. The rationale behind this choice
is to consider the DNA sequence as non-periodic but 
non-random either, in so
far as it carries information. As an example, they considered 
the Fibonacci and Thue-Morse sequences. Most of the results were 
related to the former, that is generated according to the
rules $q_a\to q_aq_b$, $q_b\to q_a$ repeteadly applied to the initial 
seed $q_a$. 

\begin{figure}
\begin{center}
\includegraphics[width=7cm, angle=270]{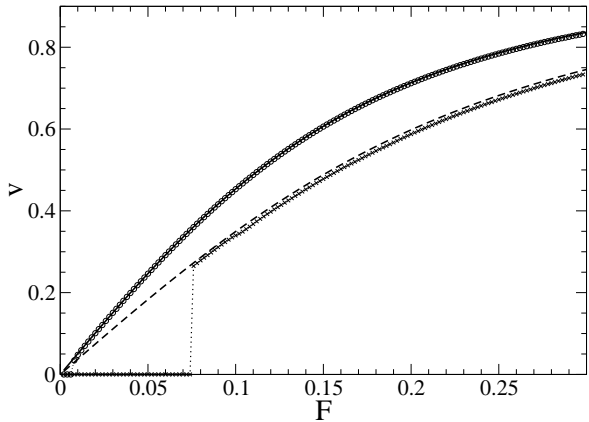}
\caption{\label{fig:23110}
Steady state velocity versus applied force for a periodic Fibonacci 
chain. The lines corresponds to
the theoretical approach and the points to numerical simulations, both for two 
periodic chains with unit cell $F_9$, of 55 sites, but with different values 
of $q_a$ and $q_b$: ($\times$) $q_a=1$ and $q_b=10$, theoretical approach
with $q_{avg}=4.436$; ($\circ$) $q_a=2$, $q_b=3$, theoretical approach
with $q_{avg}=2.382$.
}
\end{center}
\end{figure}

Being the immediate predecessor of the present work, it is worth 
summarizing here the results in \cite{yo}. 
As a first step, Dom\'\i nguez-Adame {\em et al.}\ constructed a
{\em periodic} sequence $\{q_n\}$ whose unit cell was a Fibonacci sequence 
of order $k$, $F_k$ ($F_1=q_a, F_2=q_aq_b, F_3=q_aq_bq_a, \ldots$).
This amounts to saying that the
resulting chain is given by $\{q_n\}=F_kF_kF_k...$. The number of unit
cells was chosen such that the 
length of the whole sequence $\{q_n\}$ was about 4000 sites. Subsequently,
they placed a kink of the form (\ref{ansatz}) at rest {\em in the middle 
of the sequence} and pulled it with a constant force $F$ in order to 
find the asymptotic velocity of the kink (recall that $\alpha\neq 0$ in 
Eq.\ (1), implying the existence of damping).
This computation was performed
for various orders (i.e., sizes) of the Fibonacci unit cells, $F_k$, 
up to a non-periodic, full Fibonacci sequence of 4181 sites ($k=18$).
The results of Dom\'\i nguez-Adame \emph{et al.}, 
subsequently confirmed in \cite{copia2},
pointed out the existence of an intriguing 
phenomenon, namely the existence of a threshold force for a kink 
to start moving along the chain. Once the force is above the threshold,
the kink moves asymptotically with a periodic velocity, as a consequence of
the periodicity of the constructed sequence (except in the case of the 
aperiodic $F_{18}$ sequence, in which the asymptotic, aperiodic velocity 
fluctuates around a mean value), given by the balance 
of damping and driving. Although the original derivation of
Eq.\ (\ref{vel}) is valid for a
homogeneous chain, reproducing it with a value $q_{avg}$, obtained by 
averaging $q_n$ over all lattice sites,
leads to a prediction which is very accurate, above the threshold, for 
all the periodic chains, including the final, full Fibonacci chain $F_18$.
Such prediction overestimates slightly the velocity when
increasing the ratio between the two parameters $q_a$ and
$q_b$ (see Fig.\ \ref{fig:23110} for details).
In addition, in \cite{yo},
it was also found that the threshold value depended on the length 
of the periodic chain considered: Increasing the size of the 
unit cell $F_k$ led to an increasing of the threshold, reaching a limiting 
value for the full Fibonacci sequence.

Based on the above summarized results
and on the fact that, for the same parameters,
random chains exhibited much larger threshold values (around $F\simeq 0.5$),
Dom\'\i nguez-Adame {\em et al.}\ 
concluded that long range order effects
give rise to measurable consequences on kink dynamics in aperiodic chains.
Further, they suggested that the fact that there are long range correlations
in DNA and, in any event, that it contains information and is not purely 
random, could lead to similar phenomenology in the propagation of nonlinear
coherent excitations along the molecule. At this point is where we revisit
this question,
benefiting from the huge body of genome data available nowadays.
As we will show in the next section, on the grounds of the revision 
presented here, the conjecture about the information in \cite{yo} is likely
to be incorrect, at least if we consider that the information content of 
coding DNA is different from non-coding DNA. We will come back to this 
issue in the conclusions, after discussing the results for real DNA 
sequences, in the next Section. Before proceeding to those
results, we find it interesting to extend the results in \cite{yo} by 
discussing the relevance of the choice of the initial location for the 
kink. This we do in the subsection immediately below. 

\subsection{Effect of the inital position of the kink}

\begin{figure}
\begin{center}
\includegraphics[width=7cm, angle=270]{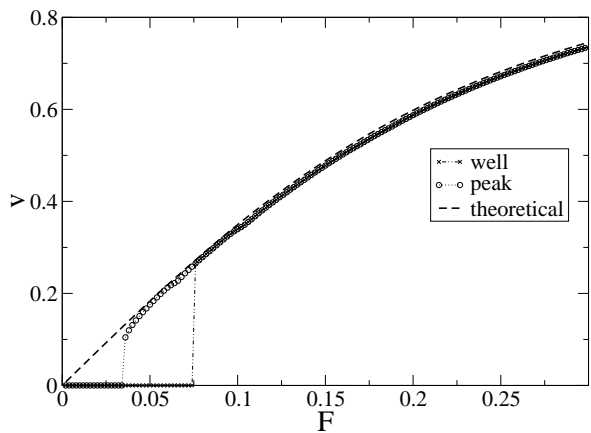}
\caption{\label{fig:picopozo}
Numerical results of steady state velocity versus applied force 
for a periodic Fibonacci chain with unit cell
$F_9$, $q_a=1$ and $q_b=10$, but starting from a peak ($\circ$) and
from a well ($\times$). The corresponding threshold force of the
simulation starting from a well is larger than the one starting from a peak.
}
\end{center}
\end{figure}

As we have said above, Dom\'\i nguez-Adame {\em et al.}\ considered 
only the existence of a threshold force for propagation on aperiodic 
chains. Although this phenomenon was addressed and interpreted correctly
in \cite{yo}, we have found that depending on the 
shape of the effective potential
around the starting position of the kink, it is more or less difficult
to make it move until the end of the lattice. If the initial position 
of a kink is a peak of the effective potential, it will cost less force to overcome
the next barrier than if the starting point is a well, as shown in Fig.~%
\ref{fig:picopozo}.

\begin{figure}
\begin{center}
\includegraphics[height=7cm, angle=270]{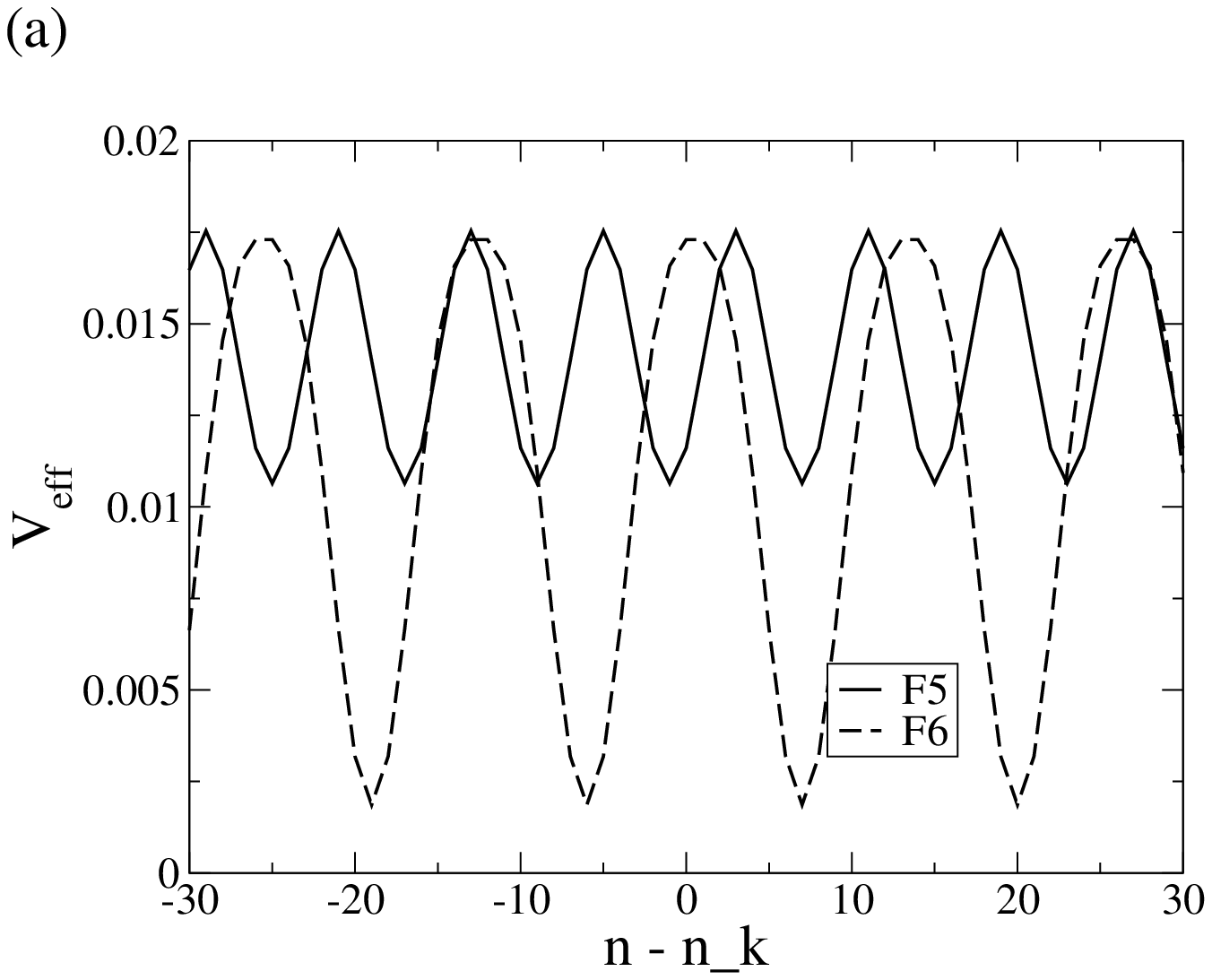}
\includegraphics[height=7cm, angle=270]{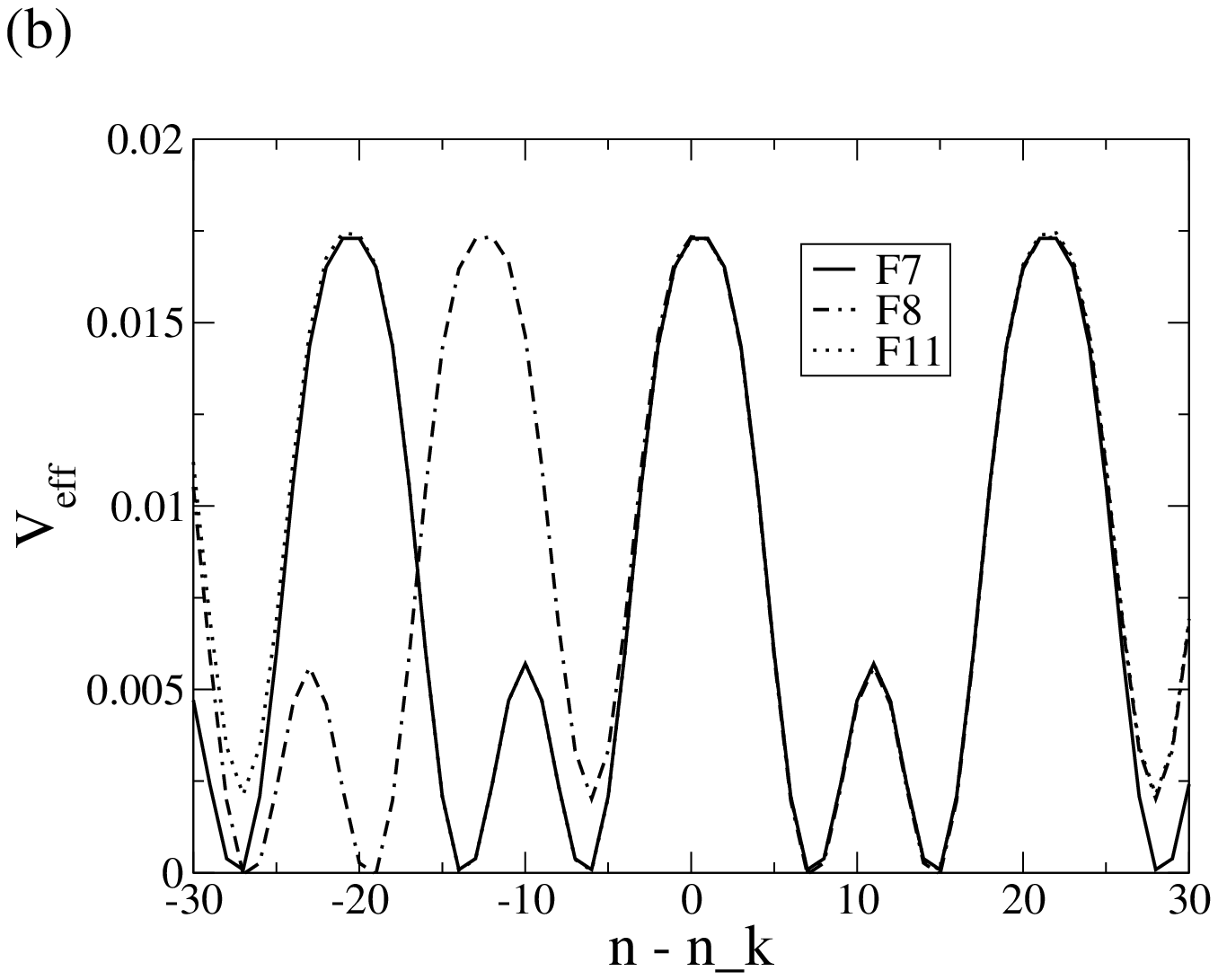}
\includegraphics[height=7cm, angle=270]{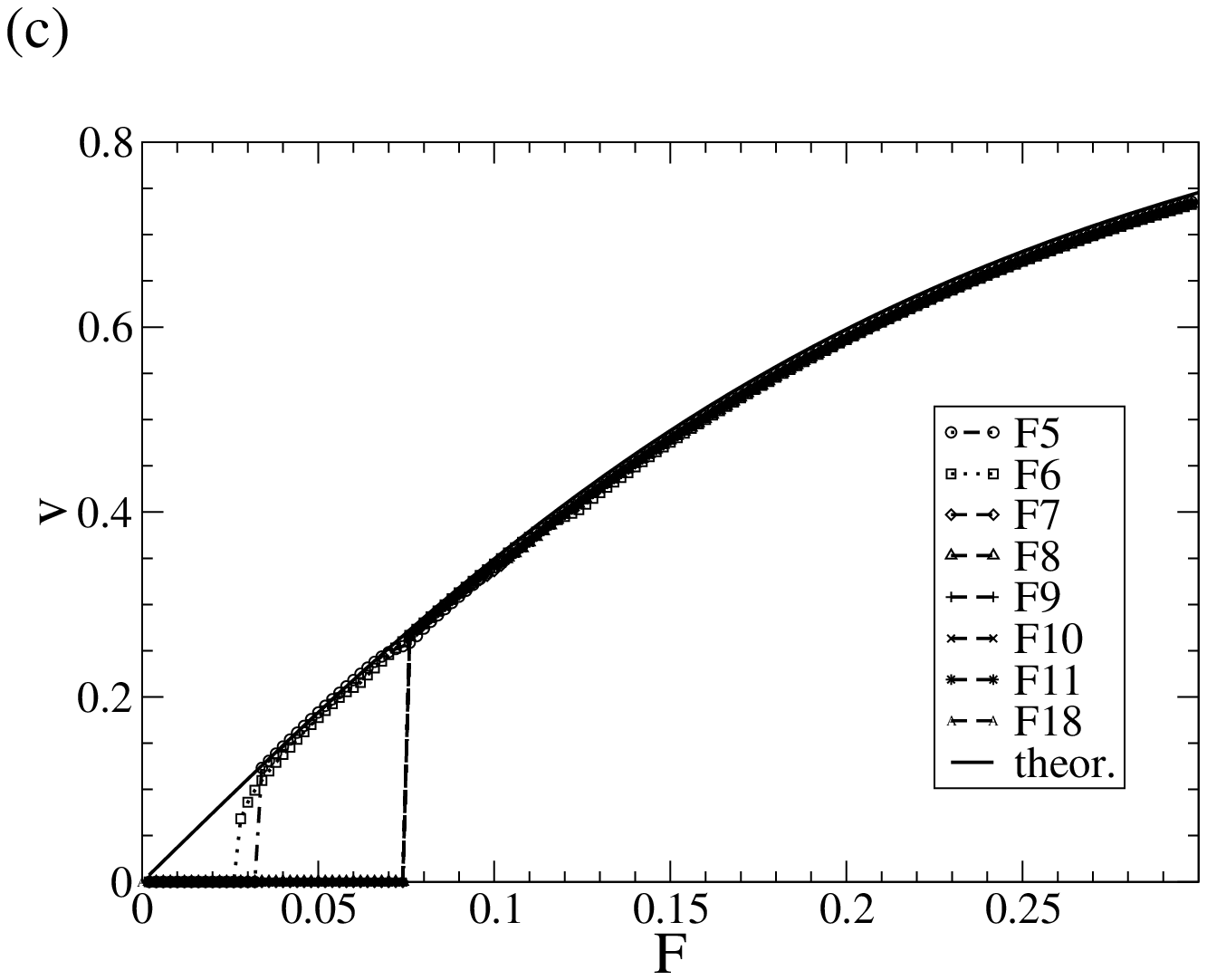}
\caption{\label{fig:Fibo}
(a) and (b) Effective potential around the beginning of a unit cell. There is 
a traslation $n_k\equiv|F_k|$ in $n$ for each sequence constructed with $F_k$'s 
in order to see the shape of the curve in the surroundings
of the beginning of a unit cell within each type of sequence. There is
another vertical shift to put the maxima of the different curves at the
same level. For clarity, the plots for $k=5, 6$ are shown in (a), 
and those for $k=7, 8, 11$ are shown in (b).
(c) Steady state mean velocity versus applied force for chains with 
different orderings and parameters $q_a=1$ and $q_b=10$. The line corresponds
to theoretical asymptotic velocity for $q_{avg}=4.438$, wich is 
the averaged value of an infinite Fibonacci sequence, $F_\infty$.
The points correspond to the numerical results for different periodic chains
with unit cell $F_k$.
}
\end{center}
\end{figure}

Given the above considerations, one can expect that for sequences formed with unit
cells of different sizes, but with identical effective potential in the 
vicinities of the starting points, the threshold force will be the same
in both cases. This can be seen in Fig.\ \ref{fig:Fibo}: In Figs.\ 
\ref{fig:Fibo} (a) and (b), we see that
for small unit cells, the
main structure of peaks and wells of the Fibonacci sequence has not been
formed yet. This can be seen, for instance, in the 
effective potentials of sequences with unit cell given by $F_7$
(of size 21) and larger, which have the same values
in the surroundings of the beginning of each unit cell. If we choose the
initial position of a kink to be near the beginning of a unit cell, 
the force needed to overcome the first barrier will be the same as
in another sequence with different unit cell, if the initial positions,
related to the beginning of each unit cell, are the same. For
unit cells of smaller size the effective pottential is different and,
therefore, so is the threshold force [see Fig. \ref{fig:Fibo} (c)]. 

\section{Effects of the sequence}
\label{sec:eff}

We now turn to the question we posed above,
namely the effects of inhomogeneity arising from the genetic sequence
on the kink propagation. To this end,
we have simulated numerically
Eq.\ (\ref{1}) with $\{q_n\}$ sequences obtained from 
the human genome. We want to stress that this description of DNA,  
arising from the original model of Englander and coworkers \cite{Englander},
intends to be only qualitatively correct. Therefore, the parameters can 
be freely chosen, trying, of course, to mimic the real ones. Accordingly,
we choose $q_a=2$ and $q_b=3$ depending on whether we have at site 
$n$ 
an A-T pair, linked by two hydrogen bonds, or a C-G pair, with three 
hydrogen bonds. The other parameters are chosen as in \cite{yo},
namely lattice spacing $a=0.1$ and damping $\alpha=0.1$. With respect to 
the lattice spacing, we want to point out that the chosen value leads to
a width of the kink which is comparable to that of spontaneous openings
of real DNA chains (some 20 to 40 base pairs).
The numerical simulations have been carried out with a standard 4th 
order Runge-Kutta scheme, with free or periodic boundary conditions. 
Finally, for the sequences we use data obtained from the
National Center for Biotechnology Information for the human genome \cite{ncbi}.

\begin{figure}
\begin{center}
\includegraphics[height=8.5cm,angle=270]{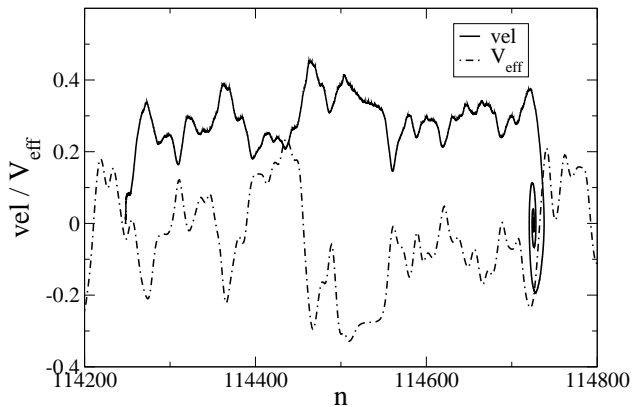}
\includegraphics[height=8.5cm,angle=270]{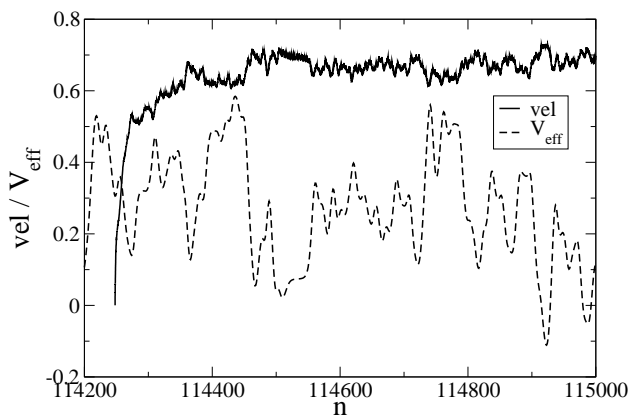}
\caption{\label{fig:velo}
Simulations of the kink soliton in the sine-Gordon model with genome data.
Shown are the velocity (solid line) and the (properly scaled to fit in the plot) 
effective potential (dashed line) 
vs the position along the chain. Up: $F=0.06$;
the kink travels along the sequence but stops at an effective potential
well.  Down: $F=0.07$; the kink travels along the whole chain.
The DNA sequence corresponds to 
contig NT\_028395.2 of human cromosome 22, between positions 114\,100 
and 115\,100, part of a gen.
}
\end{center}
\end{figure}

Typical results from our simulations are plotted in Fig.\ \ref{fig:velo},
that 
presents a phase space plot for the center of the kink
on an actual DNA sequence when 
the kink is forced 
with two different drivings. For reference, we plot in the same graph
the effective potential, that can be computed for any sequence
as described in the previous section. 
It is important to note that the effective potential does not contain 
the contribution of the force $F$, and therefore its most relevant 
information is the position of the peaks and valleys. With this in 
mind, we see that the dynamics occurs basically as in the case of the 
Fibonacci chain. There is a threshold for the kink to propagate along
the whole chain, shown by the fact that the kink ends up trapped at 
a potential well in the upper plot of Fig.\ \ref{fig:velo},
although the effective potential structure
is very different from the Fibonacci case. When the applied force exceeds 
the threshold, the kink can propagate along the whole chain (lower plot),
with a velocity which is again a (fluctuating) balance of damping and forcing. 

Figure \ref{fig:for} addresses one of the main issues we want to discuss:
the relevance of the information content for the existence and characteristics
of the threshold force. To this end, it collects our observations regarding the 
existence of thresholds in two examples:
\begin{figure}
\begin{center}
\includegraphics[height=8.5cm,angle=270]{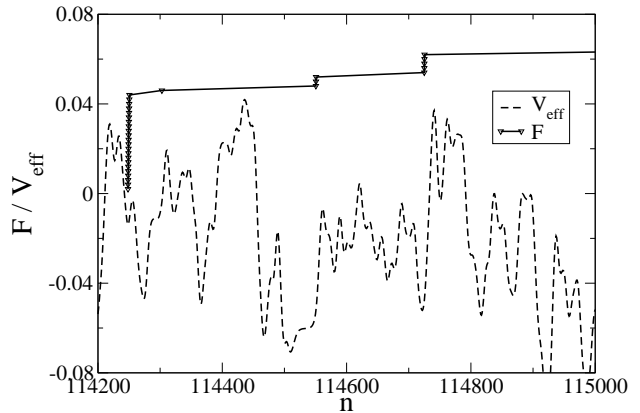}
\includegraphics[height=8.5cm,angle=270]{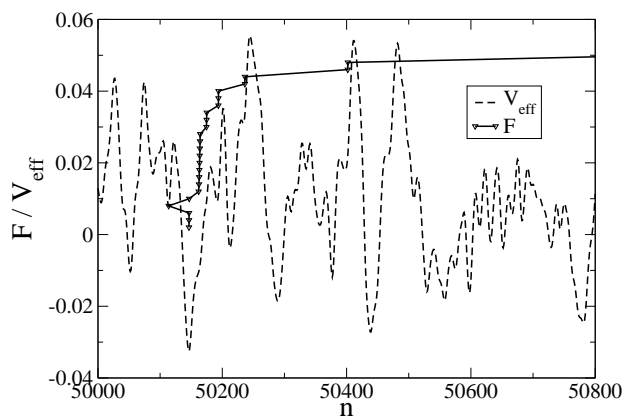}
\caption{\label{fig:for}
Simulations of the kink soliton in the sine-Gordon model with genome data.
Shown are the force (solid line) required to reach a certain position and the 
(properly scaled to fit in the plot) effective potential (dashed
line) vs the 
position along the chain. Thresholds 
can be seen as the maximum force values in each plot. 
Comparison of propagation along a coding region (up, same contig as in Fig.\ 
\ref{fig:velo}, positions 114\,000 to 115\,000) and a non-coding region
(down, same contig as in Fig.\
\ref{fig:velo}, positions 50\,000 to 51\,000).
}
\end{center}
\end{figure}
a coding region and a non-coding region. We again see the existence of 
threshold forces, in agreement with a description in terms of an 
effective potential. However, we do not observe any qualitative or 
otherwise relevant difference between the kink dynamics
(or the effective potential) in the two 
regions. This is the case with all the regions we have analyzed. 
Therefore, the conjecture by Dom\'\i nguez-Adame {\em et al.} 
based on their Fibonacci results that information may 
lead to different kink dynamical properties is, at least at the level 
of our simple model, not in agreement with the simulation results.
Of course, this is one example; although similar results are obtained 
for different coding and non-coding sequences, we can not discard that
the threshold behavior is statistically different in the two 
types of regions. However, from our simulations we consider this 
possibility quite unlikely. 

\begin{figure}
\begin{center}
\includegraphics[width=7cm,angle=270]{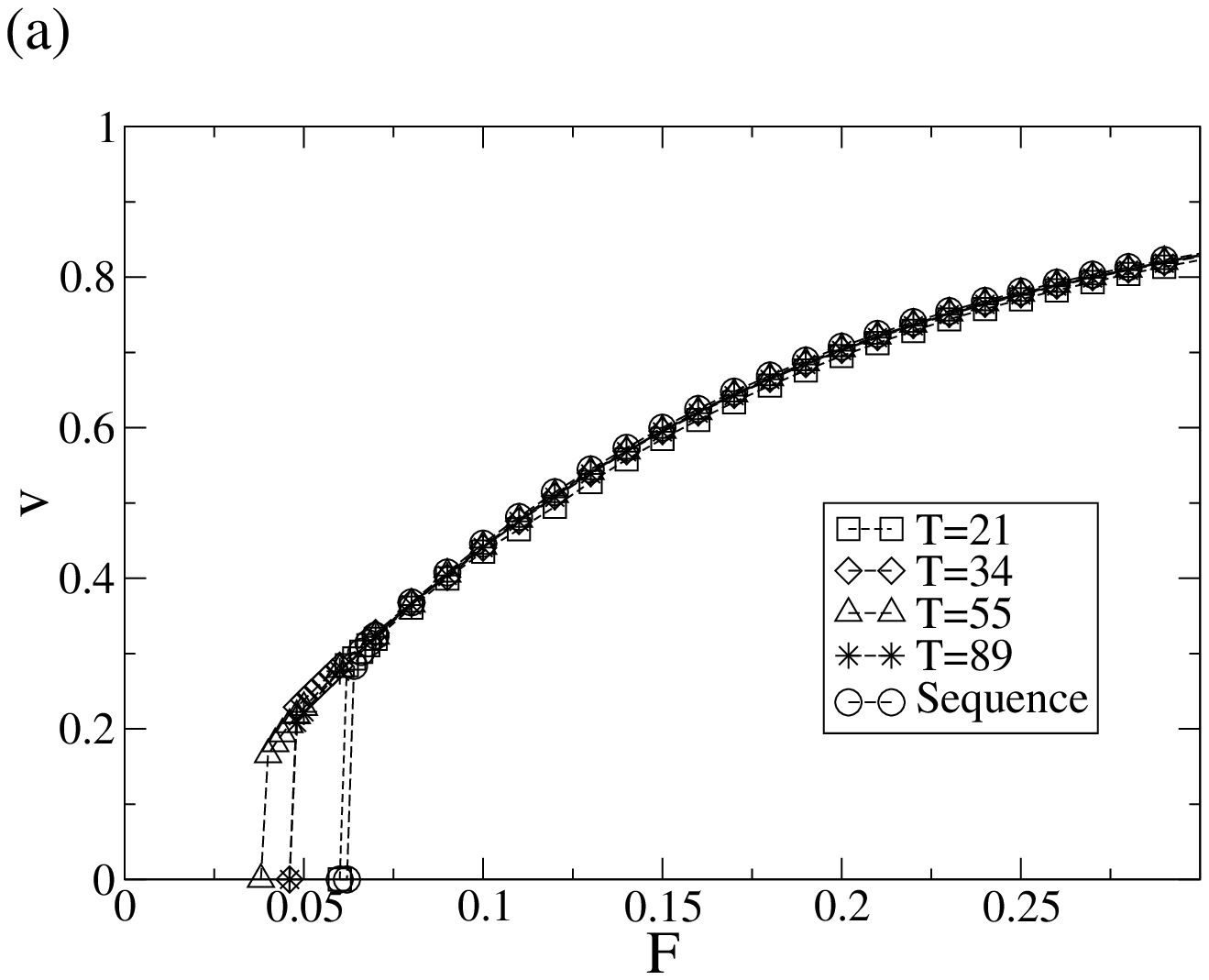}
\includegraphics[width=7cm,angle=270]{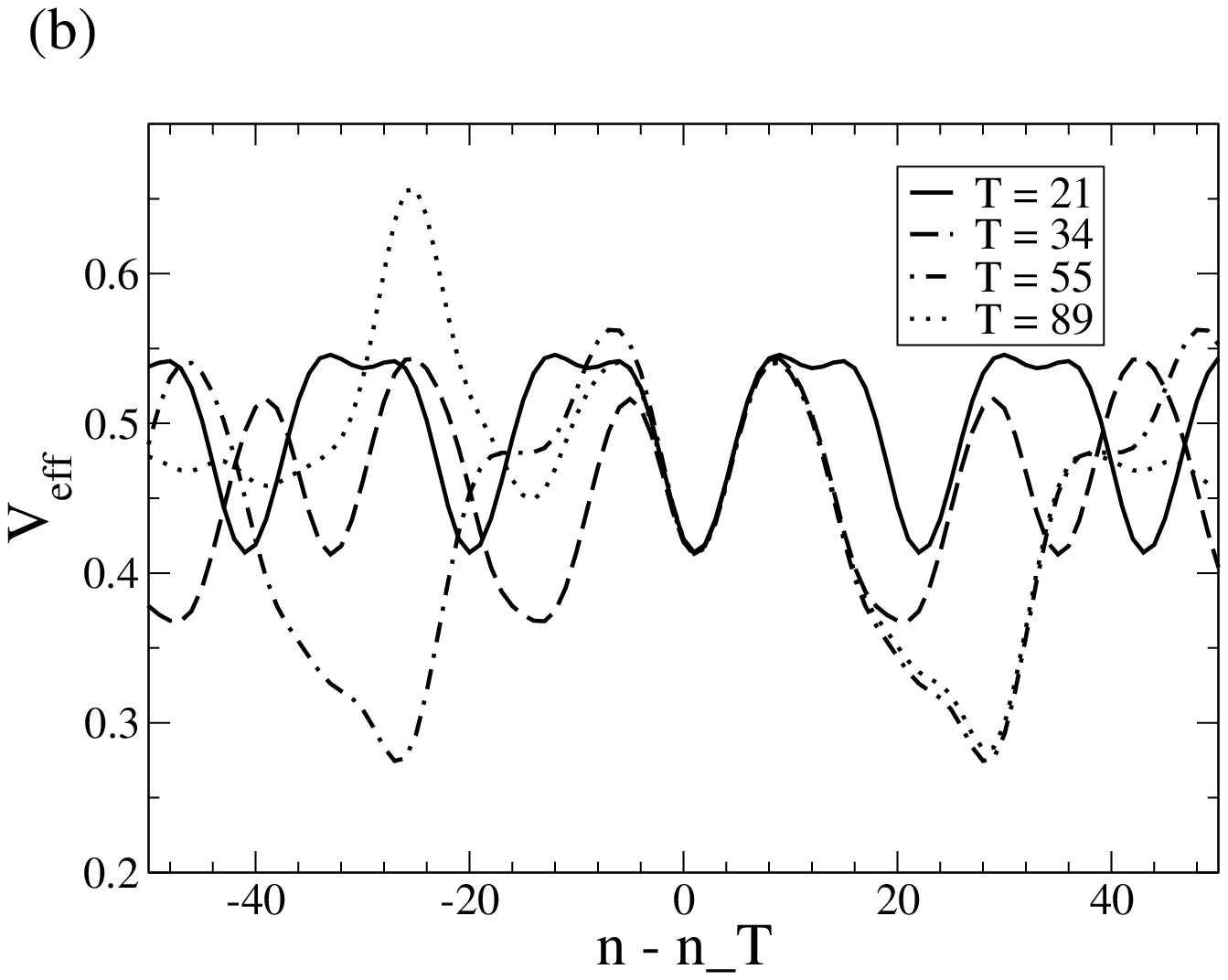}
\caption{\label{fig:seq}
(a) Velocity vs applied force for different periodic systems 
formed by repeating as a unit cell $n$ sites of the 
same contig as in Fig.\ \ref{fig:velo} beginning at 
site 114\,241, with $n=21$, 34, 55 and 89 (as in the 
Fibonacci case) and for the whole sequence going from
114\,100 to 115\,100.
(b) Effective potential for the systems in (a).
}
\end{center}
\end{figure}

Having discarded the relevance of the genetic informacion to the
determination of the threshold force, it is interesting to 
proceed with 
the comparison between 
the previous work on the Fibonacci chain and the present results,
in order to understand what is the difference and what leads to 
different conclusions. To this end, 
we have simulated periodic systems with unit cell built from 
pieces of a genetic sequence, repeated to complete a longer 
chain, much as was done with the $F_k$ building blocks in \cite{yo}
(see above). The sizes for the unit cells were chosen to mimic the 
sizes for the Fibonacci iterations, and again the starting points were
chosen in the same position related to the beginning of a unit cell. 
The results are collected in Fig.\ \ref{fig:seq} (a).
{}From this plot, we see that the most important difference between
the two cases is that in the actual DNA chain, the threshold value
does change much more for sequences with different unit cells than
in the Fibonacci case, and there is not a maximum value for the
threshold force when increasing the sizes of the unit cells.

One important point
we want to stress is that the analytical approach in 
terms of collective coordinates and an effective potential is a 
very good picture of the observed phenomenology even in this, more 
realistic setting. In fact, the results in Fig.\ \ref{fig:seq} (a) can 
be correlated with the effective potentials depicted in Fig.\ \ref{fig:seq} 
(b). On the basis of this scenario, we suggest 
that the reason for the difference between the two models lies
in the larger diversity available for possible DNA chains. The fact 
that, in Fibonacci chains,
sites of type $q_b$ are always isolated and that more than two sites of type
$q_a$ cannot be found together is very restrictive 
as to the shapes and sizes of the peaks of wells of the potential
felt by the kinks, giving rise to the Fibonacci structure mentioned
in section \ref{sec}, and therefore the Fibonacci model cannot capture
the richness of the DNA model. This is clearly seen when comparing the 
effective potential for the Fibonacci chains in Figs.\ \ref{fig:Fibo} (a) 
and (b) with that for the DNA chain in Fig.\ \ref{fig:seq} (b). 
On the other hand, when joining two unit cells in the
DNA model, the local A-T/C-G concentration can vary more than
in the Fibonacci case (where it is restricted to be close to 
a golden mean ratio), so a lower well or a higher peak may arise at the 
joining places. Finally, it is clear that considering longer and longer
sequences of DNA increases the probability of finding longer repeated
GC sequences, and hence of finding larger thresholds, as discussed 
above. 

\section{Conclusions}
\label{sec:con}

In this paper, we have reported the first conclusions of our work on the 
effects of the genetic sequence on the propagation of nonlinear excitations
along (simple models of) DNA chains. The present stage of the 
research was motivated by previous 
results on chains given by aperiodic sequences \cite{yo}, which led to 
the hypothesis that the information content of the inhomogeneities might 
be relevant for the dynamical behavior. While this general statement cannot
be ruled out at this stage (at least from a probabilistic approach as 
discussed above), we believe that at the level of the Englander 
model \cite{Englander} the simulations presented here do not confirm that
hypothesis. Specifically, we have not observed qualitative difference in
the dynamics of kinks propagating along coding or non-coding regions,
which in principle prevents the use of this simple 
model as a genome sequencing tool. 
In this respect, it is important to realize that unzipping experiments can 
not distinguish between the breaking of single base pairs, as discussed in 
\cite{ReviewMichel}. Therefore, the fact that kink pulling experiments or
simulations do not discriminate coding from non-coding region would have
a physical basis, because such discrimination would require a much better
resolution. On the other hand, we are becoming increasingly aware
that the function and information
content of non-coding regions are far from well know, and it may well be 
that the dynamics in the two types of sequences is the same because the
quantity of information is similar. Indeed, we are now beginning to realize
that the non-coding regions do have crucial implications in the DNA-protein
connections, either by inhibiting the fabrications of proteins from DNA or
by switching on and off specific genes \cite{review}. 
This is certainly an interesting issue
that deserves further attention. In this regard, it would be interesting 
to come back to the model with artificially designed sequences, with 
different information content (in the Shannon-Kolmogorov sense), in 
order to settle definitely this question. 

Even if the answer to the relation between behavior under applied forces
and information content is finally negative, there is another 
conclusion arising from our research that can be relevant from the
genomical viewpoint. We have verified that the  
effective potential approach \cite{Salkiv,yo,copia2}
to the dynamics of nonlinear excitations in DNA gives a very good 
description of the important points of the sequence, namely valleys 
and peaks. Although the precise 
dynamics of kinks is difficult to predict, and specifically, the identification
of the relevant valleys among all present in the sequence is not yet 
understood, this analytical approximation 
allows to identify the possible stop positions as well as the barriers 
that control the total opening (mechanical denaturation) of the chain. 
In fact, the complicated structure of the effective potential is 
consistent with the mechanical unzipping dynamics as observed in 
experiments \cite{Heslot}, characterized by a stick-slip behavior
with large fluctuations in the velocity (also reproduced by our 
simple model). On the other hand, the comparison with the Fibonacci 
system suggests that the features of the effective 
potential that are most relevant for the dynamics are the highest 
peaks and the deepest valleys, the total potential landscape being
responsible only for finer details. 
This is also true in the presence of thermal noise: 
Results from Langevin simulations show that 
very (in the scale of the applied forces) large temperatures do not
change much the threshold values \cite{fnl}.
This suggests that the effective potential may be a way to identify 
promoter sequences much simpler than the full simulations reported in 
\cite{copia}. In addition, the effective potential can also be used
as a tool to include in the gene identification procedure of 
Yeramian \cite{Yeramian} that may lead to results of similar quality
with much less computational effort. 
Of course, a detailed comparison with more complex models, such 
as the Peyrard-Bishop one \cite{PB,DPB,DP} is needed to ascertain 
the usefulness of the effective potential picture. Work along these 
lines is in progress. 

\acknowledgments     
 
We thank Michel Peyrard for many discussions about this work and for his 
patient explanations on DNA dynamics. We thank also Fernando Falo for 
discussions on the unzipping experiments.
This work has been
supported by the Ministerio de Ciencia y Tecnolog\'\i a of Spain
through grant BFM2003-07749-C05-01. S.C. is supported by 
a fellowship from the Consejer\'\i a de Edu\-ca\-ci\'on de la
Comunidad Aut\'onoma de Madrid and the Fondo Social Europeo.



\begin{thebibliography}{88}
\bibitem{Scott} A.\ C.\ Scott, {\em Nonlinear Science} (Oxford University,
Oxford, 1999).
\bibitem{FPU} E.\ Fermi, J.\ R.\ Pasta and S.\ Ulam, Los Alamos Report 
LA-UR-1940 (1955); reprinted in {\em Collected Papers of Enrico Fermi},
edited by E.\ Segr\'e (University of Chicago, Chicago, 1965).
\bibitem{Davydov} A.\ S.\ Davydov, Phys.\ Scripta {\bf 20}, 387 (1979).
\bibitem{Davydov2} A.\ S.\ Davydov, {\em Solitons in Molecular Systems} (2nd 
edition, Kluwer, Dordrecht, 1991).
\bibitem{Yak1} L.\ V.\ Yakushevich, Q.\ Rev.\ Biophys.\ {\bf 26}, 201 (1993). 
\bibitem{Gaeta1} G.\ Gaeta, C.\ Reiss, M.\ Peyrard and T.\ Dauxois, 
Riv.\ Nuovo Cimento {\bf 17}, 1 (1994).
\bibitem{Yak2} L.\ V.\ Yakushevich, {\em Nonlinear Physics of DNA} 
(Wiley, Chichester, 1998).
\bibitem{Gaeta2} G.\ Gaeta, J.\ Biol.\ Phys.\ {\bf 24}, 81 (1999).
\bibitem{ReviewMichel} M.\ Peyrard, Nonlinearity {\bf 17}, R1 (2004)
\bibitem{Englander} S.\ W.\ Englander, N.\ R.\ Kallenbach, A.\ J.\ Heeger, 
J.\ A.\ Krumhansl, and A.\ Litwin, Proc.\ Natl.\ Acad.\ Sci.\ USA  {\bf 77},
7222 (1980).
\bibitem{PB} M.\ Peyrard and A.\ R.\ Bishop, Phys.\ Rev.\ Lett.\ {\bf 62}, 
2755 (1989).
\bibitem{DPB} T.\ Dauxois, M.\ Peyrard, and A.\ R.\ Bishop, Phys.\ Rev.\
E {\bf 47}, R44 (1993).
\bibitem{DP} T.\ Dauxois and M.\ Peyrard, Phys.\ Rev.\
E {\bf 51}, 4027 (1995).
\bibitem{Campa} A.\ Campa and A.\ Giansanti, Phys.\ Rev.\ E {\bf 58}, 3585
(1998).
\bibitem{Wartell} R.\ M.\ Wartell and A.\ S.\ Benight, Phys.\ Rep.\ {\bf 126}, 
67 (1985).
\bibitem{Nature} A good summary of the experimental advances can be found in 
the collection of articles {\em The Double Helix --- 50 Years}, Nature 
{\bf 421}, 396 (2003).
\bibitem{Yeramian} E.\ Yeramian, Gene {\bf 255}, 139; {\em \'\i bid.}, 151; 
E.\ Yeramian, S.\ Bonnefoy and G.\ Langsley, Bioinformatics {\bf 18}, 
190; E.\ Yeramian and L.\ Jones, Nuc.\ Acids Res.\ {\b f 31}, 2843 (2003).
\bibitem{LosAlamos} G.\ Kalosakas, K.\ \O.\ Rasmussen, A.\ R.\ Bishop,
C.\ H.\ Choi, and A.\ Usheva, {\tt arXiv:cond-mat/0309157} (2003); 
C.\ H.\ Choi, G.\ Kalosakas, K.\ \O.\ Rasmussen, M.\ Hiromura, 
A.\ R.\ Bishop, and A.\ Usheva, Nuc.\ Acids Res.\ {\bf 32}, 1584--1590 (2004). 
\bibitem{yo} F.\ Dom\'\i nguez-Adame, A.\ S\'anchez, and Yu.\ S.\ Kivshar, 
Phys.\ Rev.\ E {\bf 52}, R2183 (1995). 
\bibitem{understanding} 
C.\ R.\ Calladine and H.\ R.\ Drew,
{\em Understanding DNA} (2nd edition, Academic Press, San Diego,
1997).
\bibitem{McL} D.\ W.\ McLaughlin and A.\ C.\ Scott, Phys.\ Rev.\ A {\bf 18},
1652 (1978).
\bibitem{Heslot} U.\ Bockelmann, B.\ Essevaz-Roulet y F.\ Heslot,
Phys.\ Rev.\ Lett.\ {\bf 79}, 4489 (1997).
\bibitem{Salerno} M.\ Salerno, Phys.\ Rev.\ A {\bf 44}, 5292 (1991).
\bibitem{Salerno2} M.\ Salerno, Phys.\ Lett.\ A {\bf 167}, 49 (1992).
\bibitem{copia} E.\ Lennholm and M.\ H\"ornquist, Physica D {\bf 177},
233 (2003).
\bibitem{Salkiv} M.\ Salerno and Yu.\ S.\ Kivshar, Phys.\ Lett.\ A {\bf 193},
263 (1994).
\bibitem{SB} A.\ S\'anchez and A.\ R.\ Bishop, SIAM Rev.\ \textbf{40}, 579
(1998).
\bibitem{copia2} E.\ Lennholm and M.\ H\"ornquist, Phys.\ Rev.\ E {\bf 59},
381 (1999).
\bibitem{ncbi}
{\tt http://www.ncbi.nlm.nih.gov}.
\bibitem{review} S.\ R.\ Eddy, Nature Genetics {\bf 2}, 919 (2001).
\bibitem{fnl} S.\ Cuenda and A.\ S\'anchez, Fluct.\ Noise Lett.\ 
{\bf 4}, in press (2004)
[arXiv.org preprint {\tt q-bio/0403003}].

\end{thebibliography}
\end{document}